# Nanoscale nonequilibrium dynamics and the fluctuation dissipation relation in an aging polymer glass

Hassan Oukris and N. E. Israeloff[‡]
Dept. of Physics, Northeastern University, Boston, MA 02115

**Response functions[1] and fluctuations[2] measured locally in complex materials should equally well characterize mesoscopic-scale dynamics. The fluctuation-dissipation-relation (FDR), relates the two in equilibrium, a fact used regularly, for example, to infer mechanical properties of soft matter from the fluctuations in light scattering[3]. In slowly-evolving non-equilibrium systems, such as aging spin[4,5] and structural glasses[6,7], sheared soft matter [8], and active matter[9], a form of FDR has been proposed in which an effective temperature[10], $T_{eff}$, replaces the usual temperature, and universal behavior is found in mean-field models[4,10] and simulations[6-8,11] . Thus far, only experiments on spin-glasses[12] and liquid crystals [13] have succeeded in accessing the strong aging regime, where $T_{eff} > T$ and possible scaling behavior are expected. Here we test these ideas through measurements of local dielectric response and polarization noise in an aging structural glass, poly-vinyl-acetate . The relaxation-time spectrum, as measured by noise, is compressed, and by response, is stretched, relative to equilibrium, requiring an effective temperature with a scaling behavior similar to that of certain mean-field spin-glass models.**

The FDR expresses the equilibrium thermal fluctuations of an observable, O, in terms of available thermal energy, $k_B T$, and the linear response of that observable, to an applied field, $F^{10}$. It also relates the time-dependence of the fluctuations, found in the auto-correlation function, $C(t) = <\delta O(t'+t)\delta O(t')>_{t'}$, and the time-dependent susceptibility, $\chi(t) = O(t)/F$, where F is applied at t=0. In equilibrium, C(t) should contain the same information about the dynamics that is found in χ(t), such as the spectrum of relaxation times. This is seen clearly in plotting χ(t) vs. C(t) and obtaining a straight line, with negative slope inversely proportional to temperature[10]. Deviations from the FDR have been intensively studied theoretically in aging or driven glassy systems, both on the macro[8,10] and nano-scale[11]. Aging occurs in glassy materials which have been quenched from high to low temperature. During aging, the response functions depend both on time, t, and on the age of the system since the quench, $t_w$. But χ(t,$t_w$) and C(t,$t_w$) need not have the same dependence. This can be understood conceptually by viewing aging dynamics as hopping on a tilted energy landscape[14]: energy-lowering transitions cause more rapid de-correlation than is possible by thermal energy alone. Prominent glass-transition models[15] place fragile structural glasses[16,6] in the same universality class as mean-field p-spin models with single-step replica symmetry breaking[17]. In such models[10] and simulations of structural glasses[7], the χ(t, $t_w$) vs. C(t, $t_w$), asymptotically collapse (with increasing $t_w$) to a single scaling function χ(C). χ(C) has two distinct linear regions, one agreeing with the FDR for short times, t- $t_w$ < $t_w$, then abruptly bending to a second shallower line in the strong aging regime, when t- $t_w$ > $t_w$. This second linear regime has slope reduced by a violation factor, X, relative to equilibrium, and is described as having an effective temperature, $T_{eff}$ =T/X.

Thus far, experiments studying the breakdown of the FDR in structural and soft colloidal glasses have focused on the quasi-equilibrium regime with $t_w$ > t- $t_w$, and have given a range of results that have found strong[18,19,20], weak[21], or non-existent[22] FDR violations. The strong-aging regime has been difficult to access in these systems, due largely to instrumental and statistical challenges of measuring thermal noise at the very low frequencies required. Here, using electric-force-microscopy (EFM) techniques[23][24], we probe long-lived nano-scale polarization fluctuations and dielectric responses in poly-vinyl-acetate (PVAc) films, just below the glass transition (304 K), through AC detection of local electrostatic forces. This approach enables the probing of much smaller volumes than conventional dielectric spectroscopy.   Such volumes produce large polarization fluctuations that can be detected above instrumental background down to very low frequencies.  By scanning, we can study, in effect, many samples in parallel, which improves measurement statistics. The time-dependent polarization signal, $V_P(t)$, is measured (see Methods) to observe spontaneous noise, $\delta V_P$, or after application of a tip bias, $V_{dc}$, to produce an experimental susceptibility signal $\Delta\chi_{ex}(t)=\Delta V_P(t)/V_{dc}$.

In probing nanometer scale regions, thermal fluctuations can be seen readily in the polarization as measured locally by $V_p(t)$ [23,24]. The noise can also be used to produce spatio-temporal images of the dynamics[23,24]. In this approach, $V_p$ is repeatedly measured along a one-dimensional spatial line. See figure 1. These space-time images of the surface polarization are striking, in that they clearly show the spatio-temporal aspects of glassy dynamics, e.g. the correlations seen along the time axis are distinctly longer at the lower temperatures. The auto-correlation function $C(t) =< V_P(t)V_P(t+t')>$ should relate to the glassy part of the susceptibility, $\Delta\chi_{ex}(t)$, and should obey[23,24] an FDR of the form:

$$\Delta\chi_{ex}(t) = \frac{C_{eff}}{k_B T}[C(0) - C(t)] \quad (1)$$

where $C_{eff}$ is an effective tip capacitance. $C_{eff}$ was calculated from a sphere-plane image charge model of tip capacitance for the experiment gave $C_{eff} = 6.3\pm1.8 \times 10^{-18}$ F. By plotting $\Delta\chi_{ex}(t)$ vs. $C(t)$ Equation 1 was verified for several temperatures near $T_g$ in equilibrium. The slope of the lines were nearly identical, and required $C_{eff} = 8.45 \pm 1 \times 10^{-18}$ F to produce correct temperatures reasonable agreement given the simplification of the tip geometry.

Four micron space-time images like those in fig. 1 were used to determine spatially averaged $C(t)$ and $\Delta\chi_{exp}(t)$. The instrumental resolution set by the tip radius, tip height, scan speed, and filter settings used, produced noise which was spatially correlated over 120 nm. Thus in these scans 32 independent regions were measured in parallel, and several images were averaged. For $C(t)$ scans with $V_{dc}$=constant (near 0) were used. Right and left 2s/line scans were averaged to give a 4 s sampling rate. In figure 2a, normalized $C(t)$ and $\Delta\chi_{ex}(t)$ are shown for equilibrium at 303.5 K, (~$T_g$). The shapes of the two curves are nearly identical and they are fit by stretched-exponential or Kohlrausch-Williams-Watts (KWW) functions $C(t) = C(0)\exp[-(t/\tau)^\beta]$ and $\Delta\chi_{ex}(t) = \Delta\chi_{ex}(\infty)\{1 - \exp[-(t/\tau)^\beta]\}$, where $\tau$ is the alpha-relaxation-time, and $\beta$ is a stretching exponent. For equilibrium, identical KWW parameters can be used, in this case $\beta$=0.53 and $\tau$=75 s, which confirms the FDR is satisfied. The technique produces a slightly reduced $\beta$ relative to bulk values[25].

For aging experiments, the samples were heated to well above $T_g$ to 324 K, held for 1 minute, cooled and stabilized at final temperature, $T_F$, with a cooling rate of 5K/min through $T_g$. (See temperature profile in fig. 2b). For $T_F$ = 302K, where the equilibrium $\tau$~150 s there were no observable FDR violations. The data and analysis shown here are for $T_F$=298K, where $\tau \sim 2500$s. The start of aging, $t_w$ =0, was defined as the dynamic $T_g$ (~304K) crossing. Topographic images showed that thermal drift of the tip position was significant up to $t_w \sim$ 3min. Thereafter, space-time images were collected, and the small drift contribution to C at early $t_w$ could be determined and subtracted. Ten or more quenches were used for each correlation measurement (> 320 samples) and six or more (>192 samples) for relaxation. For relaxation $V_{dc}$=0.2V was applied at $t_w$ to give $\Delta\chi_{ex}(t, t_w)$. Data points at $at_w$ and t were pair-wise correlated to calculate $C(t, t_w)$, but to improve statistics, $C(t, t_w)$ was averaged over 1 minute wide windows of time centered on $t_w$ and t.

In fig. 2b normalized correlation and relaxation curves are shown for aging with $t_w$= 3min. The shapes of the two curves are now very different, with $\Delta\chi_{ex}(t, t_w)$ more stretched compared with $C(t, t_w)$. A single KWW is no longer an ideal fit but can be used to fit reasonably well the intermediate to long times, with $\beta$=0.64±0.04 for C and $\beta$=0.39±0.04 for $\Delta\chi_{ex}$ with $\tau$=385 s. Aging of response in structural glasses for shallow quenches can be understood as a slow evolution from higher (fictive) temperature dynamics to the lower temperature equilibrium dynamics, modeled as an evolving relaxation rate $(1/\tau)$[26] which stretches out the dynamics and gives a reduced $\beta$. More surprising is that the correlation function has an increased $\beta$, indicating a *compressed* spectrum of relaxation rates.

Given the different stretching parameter, $\beta$, for $\chi$ and C during aging, the FDR clearly can no longer hold. In figure 3a a plot of $\Delta\chi_{ex}(t, t_w)$ vs. $C(t, t_w)$ for various $t_w$ and equilibrium is shown. While the equilibrium curve is linear, as predicted by eqn. 1, the aging curves all exhibit a distinct curvature, trending toward horizontal, indicating failure of FDR, at smallest values of C, in the strong aging regime. This is qualitatively what is predicted by various mean-field models and simulations[7,10,11]. Most striking is that there appears to be a near-collapse of all curves onto a global curve, $\chi(C)$. Such a collapse is only achieved in the asymptotic $t_w$ limit in mean-field spin-glass models[10] and with scaled axes in spin-glass experiments[12]. However in Lennard-Jones structural glass simulations, a similar collapse was found[7]. The local slope of the global $\chi(C)$ curve gives the violation factor: $X(C) = -d\chi(C)/dC$[10]. In simulations and mean-field p-spin models, and spin-glass experiments, a discontinuous step-function $X(C)$, with two distinct values, $X=1$ for small $(t-t_w)/t_w$, and $X<1$ for strong aging[6,16]. The class of models including the Sherrington-Kirkpatrick (SK) spin-glass model which exhibit continuous replica symmetry breaking, have a continuous $X(C)$.[5,17] In the present case, if we join all the of the data of fig 3a, and treat it as a single function, $\chi(C)$, we can determine $X(C)$ from the the local slope of this curve. In fig. 3b the calculated $X(C)$ is shown. This function appears to be continuous: a linear or power-law fit work equally well, with $R^2=0.80$. The power-law fit is shown, giving $X(C) \sim C^{0.57}$. A step-function fits the data rather poorly, with $R^2 = 0.35$

      We have found that during aging in a fragile structural glass, the spontaneous dynamics are very different from relaxation dynamics. The spectrum of relaxation times appears stretched in a relaxation experiment, and compressed in a fluctuation measurement, compared with the equilibrium spectrum, requiring a modified form of FDR. The way in which the equilibrium FDR is modified may tell us something about the way the glassy states are organized[6,27] in equilibrium. A single effective temperature during aging was not found, but the collapse of the data to $\chi(C)$, with an apparent continuous FDR violation factor, $X(C)$, are distinctive signatures, similar to an SK class of mean-field spin-glass models, and therefore provides guidance in finding a successful model of fragile structural glasses. With larger data sets, a fixed-t analysis could more accurately determine effective temperatures[28] and $X(C)$. Open questions include the

temperature dependence of X for deep quenches, and how $T_{eff}$ can be larger than the initial temperature of the system. The role of fragility should be explored, since in a fragile glass former, dynamics slow dramatically in a small range of temperature, requiring energy landscape changes much larger than $k_BT$.

**Methods**

A variation of SPM, electric force microscopy (EFM) is used, which involves oscillating a small silicon cantilever with sharp metal-coated tip at its resonance frequency, $f_0$, in ultra-high-vacuum. The EFM tip is held a distance, $z=18\pm5$nm, above dielectric sample, and biased with voltage, V, relative to a conducting substrate on which the sample is coated. See figure 1 inset. The bias provides an electrostatic force between the tip and sample surface, which shifts the cantilever resonance frequency by $\delta f(V)$, which is then measured. The tip electrostatic force can be calculated from the charging energy on the tip capacitance, $U =1/2\ C_{tip}V^2$ and $F=-dU/dz$. The bias induced shift in cantilever resonance frequency, $\delta f(V)$, can be obtained from the force gradient $dF/dz$ which supplies a fractional change in the cantilever spring constant, k. Sample polarization produces a surface potential and an image charge on the tip, both acting to produce an effective dc offset, $V_P$, to the applied bias. The electrostatic component of $\delta f$ is therefore:

$$\delta f = \frac{1}{2}\frac{f_0}{k}\frac{\partial F}{\partial z} = -\frac{1}{4}\frac{f_0}{k}\frac{\partial^2 C_{tip}}{\partial z^2}(V-V_P)^2$$

Because the signal is proportional to force derivative on the sharp tip, it is most sensitive to sample polarization near the surface. For a conical tip with typical 30 nm tip radius, a region about 60 nm in diameter and 20 nm in depth below the surface is probed[25].

By applying a 200 Hz oscillating bias, $V_{ac}= V_0\sin(\omega t)$, with $V_0 =1$ V and using a lock-in amplifier to detect the oscillation in $\delta f \sim 2V_0V_p\sin\omega t$, the polarization contribution to the tip-sample interaction can be isolated, and $V_p$ can determined. To study dielectric relaxation a dc bias, $V_{dc}=0.2$V is added as a step function to the ac bias, the resulting time-dependent response of $V_p(t)$ is measured, and an experimental susceptibility is defined as $\Delta\chi_{ex}(t)=\Delta V_P(t)/V_{dc}$, where $\Delta V_P(t)$ is the slow portion of the response. Thus is $\Delta\chi_{ex}(t)$ is proportional to the glassy part of the dielectric susceptibility, $\Delta\chi(t)$. The samples studied here are poly-vinyl-acetate (PVAc), $M_w=167000$, (Sigma) which has a bulk glass transition of $T_g= 308$ K. Thin films (1 micron) of PVAc were prepared by dissolving the polymer in toluene and spinning the solution onto Au-coated glass substrates, drying in air and annealing for 24 h in high vacuum at $T_g$. Samples were mounted in a UHV SPM system (RHK SPM 350) for measurements onto a chilled-water cooled stage. Samples are radiatively heated from below and temperature was measured with a small thermocouple clamped to the sample surface. Previously we showed that the near-surface $T_g$ of PVAc is mildly suppressed by ~ 5 K[25], as we see in the current experiment.

For the FDR analysis, we use the standard method of plotting fixed-$t_w$ curves. Although this has been shown[28] to underestimate $T_{eff}$, an attempt to use the more accurate fixed-t analysis[29], produces very poor quality curves. Both methods produce the two slope behavior in Lennard-Jones glasses[28].

‡ Author to whom correspondence and requests for materials should be addressed

**Acknowledgements**

We acknowledge the support of NSF grant DMR 00606090. We thank Leticia Cugliandolo for helpful discussions.

**Contributions**

HO- experimental work, data analysis, writing paper

NEI-project planning, experimental work, data analysis, writing paper

**Figures**

Figure 1. Spatio-temporal images of thermal fluctuations measured on a single spatial dimension in equilibrium. Gray scale indicates sample polarization. Top: 301.5 K, showing long-lasting correlations relative to the bottom image at 305.5 K. Inset: measurement set-up showing EFM cantilver with conducting tip, biased with a sinusoidal voltage, above a dielectric polymer sample coated on a conducting electrode.

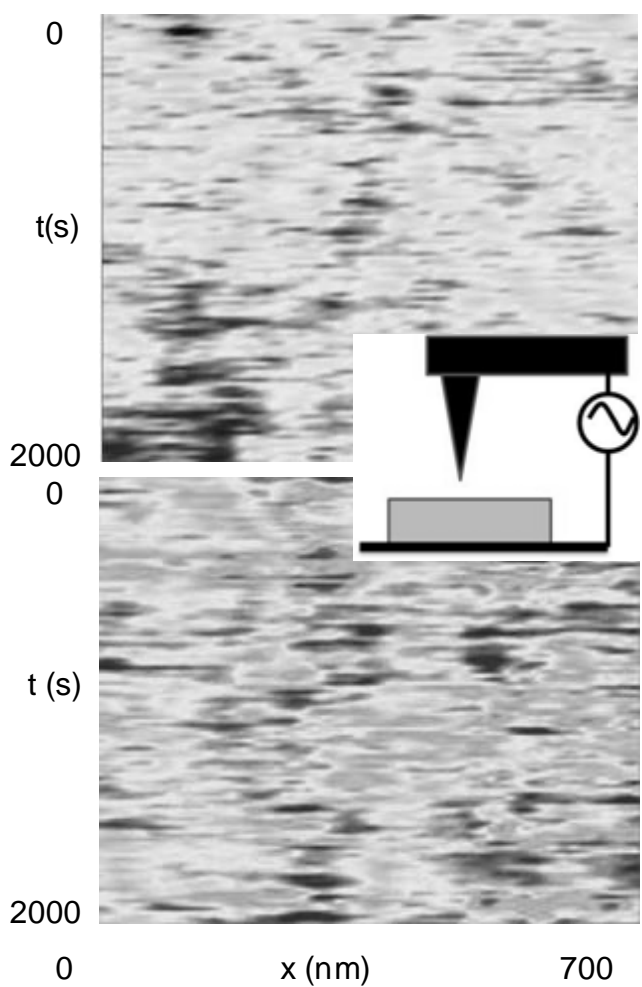

Figure 2 Normalized local dielectric response ($\chi$) and correlation (C) functions measured in equilibrium (a) and during aging (b), all with stretched-exponential (KWW) fits. Equlibrium $\chi$ and C curves are fit with identical KWW parameters: $\tau$=75 s and $\beta$=0.53. The aging curves in (b) require different stretching exponents. For $\chi$, $\beta$=0.40±0.04, and for C, $\beta$=0.65±0.04, and both require $\tau$ =385 s. Inset: temperature quench profile with $T_g$ indicated as the start of aging.

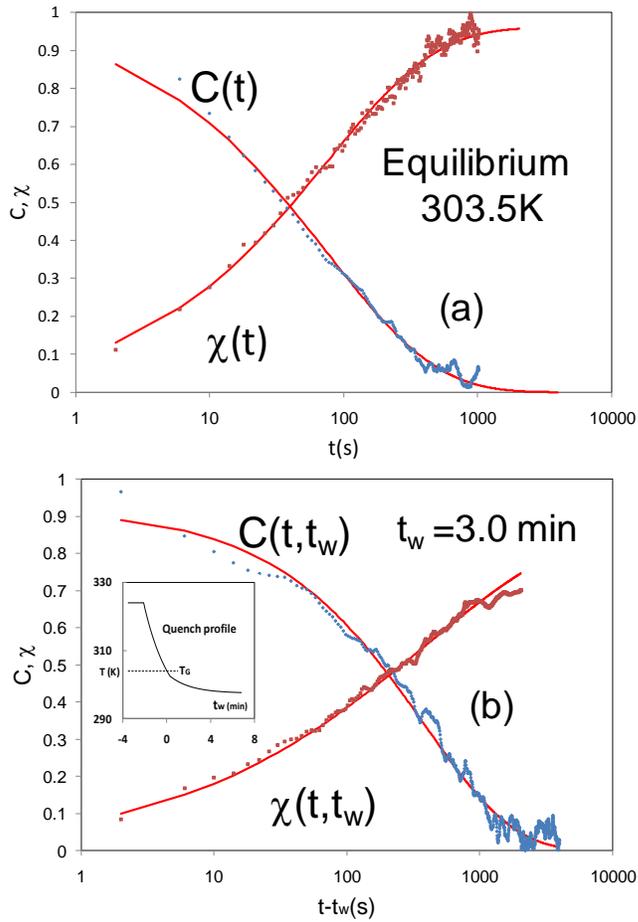

Figure 3. Susceptibility vs. correlation for aging and equilibrium. In (a) normalized $\chi(t,t_w)$ is plotted vs. normalized $C(t,t_w)$ for various waiting times after the quench. Equlibrium data for 303K is also plotted along with its linear fit. In (b) FDR violation factor, X, of all aging data in (a) joined and treated as a single universal curve, is plotted on a log-log plot vs. C, and fit to a power-law.

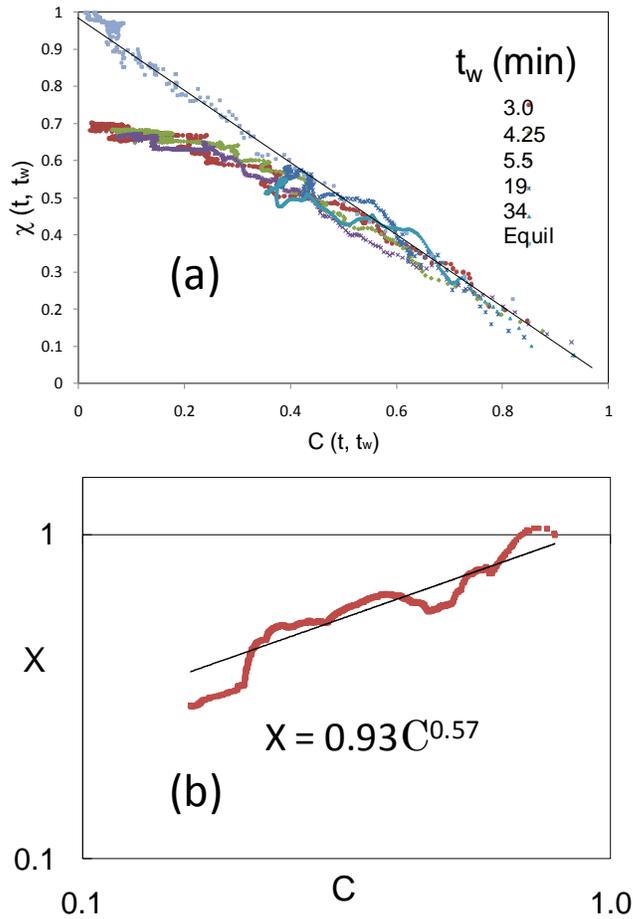